\documentclass[aip,jcp,preprint]{revtex4-1}
\usepackage{colordvi,epsfig,color,amsmath,amssymb}

\begin{document}

\def\be{\begin{equation}}
\def\ee{\end{equation}}
\def\bee{\begin{eqnarray}}
\def\eee{\end{eqnarray}}
\def\sech{\mbox{sech}}
\def\e{{\rm e}}
\def\d{{\rm d}}
\def\L{{\cal L}}
\def\U{{\cal U}}
\def\M{{\cal M}}
\def\T{{\cal T}}
\def\V{{\cal V}}
\def\R{{\cal R}}
\def\kb{k_{\rm B}}
\def\tw{t_{\rm w}}
\def\ts{t_{\rm s}}
\def\Tc{T_{\rm c}}
\def\gs{\gamma_{\rm s}}
\def\tm{tunneling model }
\def\TM{tunneling model }
\def\tilde{\widetilde}
\def\Deltac{\Delta_{0\rm c}}
\def\Deltamin{\Delta_{0\rm min}}
\def\Emin{E_{\rm min}}
\def\tauc{\tau_{\rm c}}
\def\tauac{\tau_{\rm AC}}
\def\tauw{\tau_{\rm w}}
\def\taumin{\tau_{\rm min}}
\def\taumax{\tau_{\rm max}}
\def\de{\delta\varepsilon / \varepsilon}
\def\pF{{\bf pF}}
\def\pFAC{{\bf pF}_{\rm AC}}
\def\halb{\mbox{$\frac{1}{2}$}}
\def\ihalb{\mbox{$\frac{i}{2}$}}
\def\dreihalb{\mbox{$\frac{3}{2}$}}
\def\viertel{\mbox{$\frac{1}{4}$}}
\def\achtel{\mbox{$\frac{1}{8}$}}
\def\with{\quad\mbox{with}\quad}
\def\und{\quad\mbox{and}\quad}
\def\za{\sigma_z^{(1)}}
\def\zb{\sigma_z^{(2)}}
\def\ya{\sigma_y^{(1)}}
\def\yb{\sigma_y^{(2)}}
\def\xa{\sigma_x^{(1)}}
\def\xb{\sigma_x^{(2)}}
\def\spur#1{\mbox{Tr}\left\{ #1\right\}}
\def\erwart#1{\left\langle #1 \right\rangle}
\newcommand{\bbbone}{{\mathchoice {\rm 1\mskip -4mu l}{\rm 1\mskip -4mu l}{\rm
1\mskip -4.5mu l}{\rm 1\mskip -5mu l}}}

\newcommand{\veco}[2]{(#1)\hat{e}_{#2}}
\renewcommand{\v}[1]{\ensuremath{\mathbf{#1}}}
\newcommand{\mean}[1]{\langle #1 \rangle}

\title{Probing chirality fluctuations in molecules by nonlinear optical
spectroscopy}

\author{N. Mann$^{1,2}$, P.~Nalbach$^{1,2}$, S. Mukamel$^{3}$ and
M.~Thorwart$^{1,2}$}
\affiliation{
$^1$I.\ Institut f\"ur Theoretische Physik, Universit\"at Hamburg,
Jungiusstra{\ss}e 9, 20355 Hamburg, Germany\\
$^2$The Hamburg Centre for Ultrafast Imaging, Luruper Chaussee 149, 22761
Hamburg, Germany\\
$^3$Department of Chemistry, University of California, Irvine, California
92697-2025, USA}

\date{\today}

\begin{abstract}

Symmetry  breaking caused  by geometric fluctuations  can  enable  processes 
that are otherwise forbidden. An example is a perylene bisimide dyad whose
dipole moments are perpendicular to each other. F\"orster-type energy transfer
is thus forbidden at the equilibrium geometry since the dipolar coupling
vanishes. Yet, fluctuations of the geometric arrangement have been shown to
induce finite energy transfer that depends on the  dipole variance, rather than
the mean. We demonstrate an analogous effect associated with chirality 
symmetry breaking. In its equilibrium geometry this dimer is non chiral. The
linear chiral response which depends on the average geometry thus
vanishes. However, we show that certain 2D chiral optical signals are finite 
due to geometric fluctuations. Furthermore, the correlation time of these
fluctuations  can be experimentally revealed by the waiting time dependence
of the 2D signal.

\end{abstract}


\maketitle

\section{Introduction}

Almost all biological molecules such as nuclear bases, sugars and peptides are
chiral. These chiral structures occur in enantiomer pairs connected by a
reflection symmetry. While chiral enantiomers show large differences in their
biological activity and their chemical reactivity, most physical properties are
identical. Thus, only few physical methods are usable to study chirality
\cite{Chiral1,Chiral2}. Most commonly, the circular dichroism (CD)
\cite{Chiral3}, the difference between the absorption of left- and right-handed
circularly polarized light, is used. Equivalent information to that from CD can
be obtained from the $I_{yz}-I_{zy}$ tensor component of the free induction
decay signal for linearly polarized light propagating along the $x$-direction 
\cite{Chiral4}. 
Often, chiral molecules have more than a single conformation which are 
thermally accessible at finite temperatures. Each conformation has 
its own chiral response, and, thus, only a thermal average over these conformations 
can be measured in the experiment \cite{Chiral5}. 
Recently, nonlinear optical signals have been proposed as a
measure for chirality. They, in addition, allow us to determine the relevant
time scale of the chirality switching between conformations and enantiomers 
\cite{Sanda2011}.
By these nonlinear optical signals, also molecules can be studied 
which are achiral in their equilibrium configuration, but show a finite 
chirality when thermal fluctuations of their configuration break spatial
symmetry \cite{Sanda2011}. They also can reveal the correlation time of
such thermal fluctuations directly in experiments. The nonlinear optical
signals also provide useful information when they are obtained from molecules
in the bulk after averaging over all molecular orientations. They in particular 
 do not suffer from the artifacts of the fluorescence detected linear CD which
even may occur in the linear dichroism in single immobilized molecules
\cite{Cohen09}.

\begin{figure}[t]
\epsfig{file=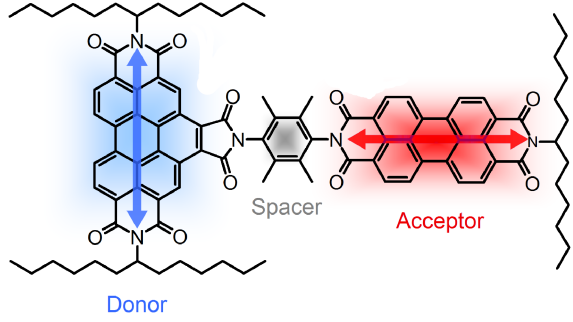,width=8cm}
\caption{\label{fig1} Sketch of the chemical structure of an orthogonally
arranged perylene bisimide donor acceptor pair (PBDA) together with 
the transition dipole moments \cite{Langhals2008}.}
\end{figure}

Here, we focus on geometry fluctuations in a perylene bisimide
donor acceptor (PBDA) pair. These have been shown to induce rather strong 
F\"orster resonant energy transfer  (FRET) that is
forbidden in the average geometry \cite{Nalbach2012}. We extend the same idea to
chirality and show that a molecule which is achiral in its equilibrium
configuration may show signatures of chirality induced by geometry fluctuations
in its nonlinear optical 2D spectrum. 

FRET \cite{Forster:AnnPhys1948} is a well established and widely used measuring
tool to determine the molecular proximity of light-absorbing and fluorescent
structures \cite{MayKuehn,vanAmerongen00}. These applications rely
on the basic property of FRET that the energy transfer time is proportional to
the dipolar coupling strength between the transition dipole moments
$\vec{{\boldsymbol \mu}}_j$ of the energy donor / acceptor ($j=1/2$) spaced at
distance $R$ with connecting unit vector $\vec{{\bf n}}$, i.e., 
$\tau_{FRET}\propto ( \left[ \vec{{\boldsymbol \mu}}_1\vec{{\boldsymbol \mu}}_2 
 - 3(\vec{{\boldsymbol \mu}}_1\vec{{\bf n}}) (\vec{{\boldsymbol \mu}}_2\vec{{\bf
n}}) \right] / R^3 )^{-2}$.
Accordingly, FRET vanishes for orthogonally arranged dipoles when
$\vec{{\boldsymbol \mu}}_1\vec{{\boldsymbol \mu}}_2 =0$ and $\vec{{\boldsymbol
\mu}}_i\vec{{\bf n}}=0$. In this work, we consider a perylene bisimide
donor acceptor pair, which is a heterodimer and has such a property. It has also been analyzed in
recent experiments 
\cite{Langhals2008,Langhals2010} and its chemical structure is sketched in Fig.\
\ref{fig1}. Surprisingly, despite the orthogonal arrangement, a fast energy
transfer was measured with a transfer time of $9.4$~ps for PBDA in
chloroform \cite{Langhals2010}. It could be explained on the basis of angular
thermal fluctuations in the geometric structure \cite{Nalbach2012}. These angle
fluctuations induce fluctuations of the dipolar energies and contribute to the
FRET \cite{Jang07,Silbey09}. Treating the angle fluctuations as environmental
fluctuations results in strong effective excitonic dipolar donor-acceptor
couplings and thus in a considerable noise-induced energy transfer.
The angular fluctuation strength can be estimated from the energy fluctuations
which are observable by the optical Stokes' shifts. Then, the energy transfer
time and its temperature dependence even quantitatively could be accounted for
\cite{Nalbach2012}. Moreover, the distance dependence of $\tau_{FRET}$ is
modified to $\propto R^3$ as compared to $\propto R^6$ from standard F\"orster
theory \cite{Nalbach2012}.   
 
The quantitative success of these results crucially depends on the relation
between energy and angular fluctuations. The angular fluctuations
are not directly accessible experimentally so far which finally prevents an
experimental verification. In equilibrium, the PBDA pair is an achiral molecular
structure. However, angular fluctuations clearly break this symmetry and 
result also in a finite chirality. This can be quantified by the nonlinear
spectroscopic signals, as it was proposed by Sanda {\sl et al.\/}
\cite{Sanda2011}. We apply this concept of chirality fluctuations to the
particular case of the PBDA pair and determine the chiral signals
by treating the slow angular fluctuations as an Ornstein-Uhlenbeck process. 
We calculate for a finite fixed angle the $I_{yz}-I_{zy}$ tensor component of
free induction decay for linearly polarized light propagating along the
$x$-direction. This yields the information which is equivalent to the circular
dichroism. We also study how the response changes when tuning the 
PBDA pair towards a homodimer. Such an arrangement in particular facilitates the  
investigation of the noise-induced energy transfer \cite{Nalbach2012}. 
Then, we study the angle-averaged 2D chiral spectrum and its
dependence on the waiting time. In particular, this allows us to determine the
angle fluctuation correlation time and strength. We show quantitatively that
the chirality fluctuations can be used to test the orthogonality of the dipoles
in the PBDA pair on an entirely independent footing. 

\section{Model}

\subsection{Geometric set-up}

\begin{figure}[t]
\epsfig{file=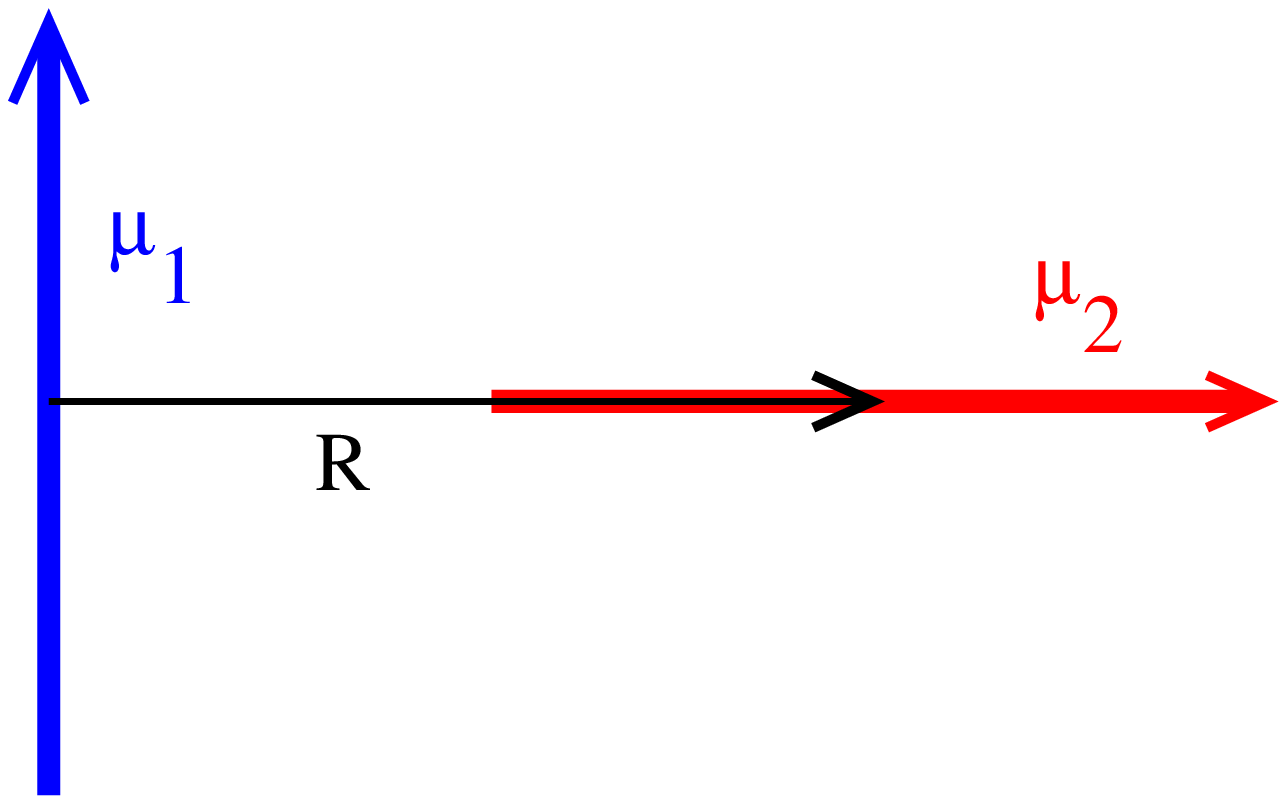,width=3.5cm}
\epsfig{file=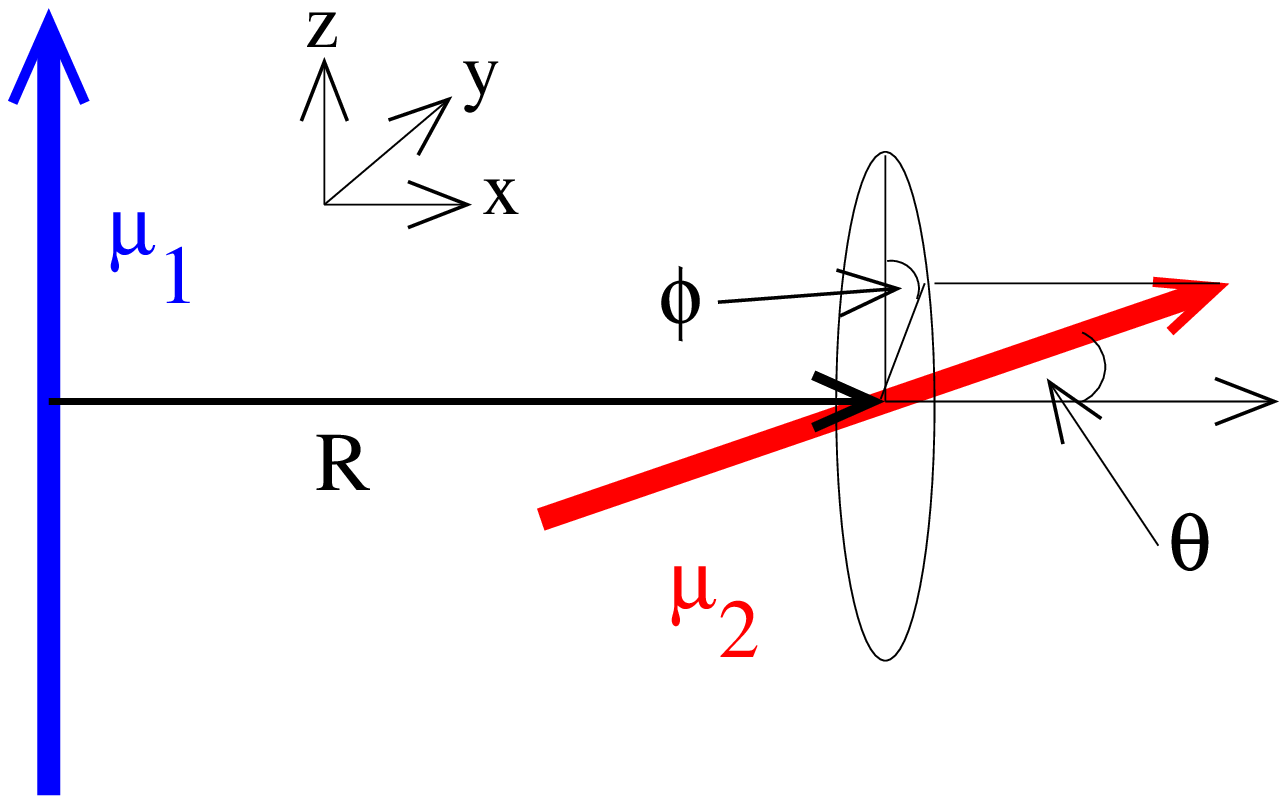,width=3.5cm}
\caption{\label{fig2} Left: Sketch of the arrangement of the two transition
dipole moments $\boldsymbol{\mu}_1$ and $\boldsymbol{\mu}_2$ and the connecting
vector $\mathbf{R}$ of the PBDA pair. Right: Illustration of the angle
$\theta$ and $\phi$ when $\boldsymbol{\mu}_2$ does not point along the
$x$-axis.}
\end{figure}

The focus of our study is an orthogonally arranged perylene bisimide donor
acceptor pair (PBDA) which is sketched in Fig.\ \ref{fig1}. To model excitonic
energy transfer, the arrangement of the electric dipoles is relevant.
Fig.\ \ref{fig2} displays the
perpendicular dipole moments $\boldsymbol{\mu}_1$ and $\boldsymbol{\mu}_2$ and
the connecting vector $\mathbf{R}$, which is parallel to  $\boldsymbol{\mu}_2$
and perpendicular to $\boldsymbol{\mu}_1$. Thus,
$\mathbf{n}\perp\boldsymbol{\mu}_1\perp\boldsymbol{\mu}_2$ and 
$\mathbf{n}\parallel\boldsymbol{\mu}_2$ and the dipole-dipole coupling strength 
\be
J =
\frac{\boldsymbol{\mu}_1\boldsymbol{\mu}_2-3(\boldsymbol{\mu}_1\mathbf{n}
)(\boldsymbol{\mu}_2\mathbf{n})}{|\mathbf{R}|^3} 
\ee
vanishes accordingly. Here, $\mathbf{n}$ is the unit vector in the direction of
$\mathbf{R}$.

Since the donor and the acceptor are rigid, any deviation from the orthogonal
arrangement of the dipoles should arise from rotations at the location
of the chemical bonds between the spacer and the donor and between the spacer
and the acceptor. In order to simplify the following calculations,  we assume
that the
connecting vector fixes the coordinate system and the two dipole moments can
only rotate around their centers. Rotations of $\boldsymbol{\mu}_1$ might then
result in a finite dipole-dipole coupling, but the PBDA is still achiral. In
other words, since $\mathbf{n}\parallel\boldsymbol{\mu}_2$ any additional vector
will span a plane with the former two vectors and the whole system is planar.
Thus, a chiral signal can only result when $\boldsymbol{\mu}_2$ rotates out of
the plane of $\boldsymbol{\mu}_1$ and $\mathbf{n}$, thereby forming a helical
structure.
For simplicity, we disregard rotations of $\boldsymbol{\mu}_1$ completely since
for not too strong angular fluctuations, they will leave our
results qualitatively unchanged. 

To describe the rotation of $\boldsymbol{\mu}_2$, we introduce two angles.
First, $\theta$ measures the angle between $\boldsymbol{\mu}_2$ and the
$x$-axis. Second, $\phi$ denotes the angle between the projection of
$\boldsymbol{\mu}_2$ in the $y-z$ plane and the $z$-axis (see Fig.\ \ref{fig2}).
 We note that rotations with $\phi=0$
cause a finite dipolar coupling although the complex remains planar and thus
achiral. For $\phi=\pi/2$, however, the complex is chiral but the dipolar
coupling vanishes. To observe an optical chiral signal from and a 
finite energy transfer in the complex, both a
finite dipolar coupling and a chiral geometry are necessary. Thus, we have to
consider variations in both angles, $\theta$ and $\phi$, in contrast to the
somewhat simpler situation of the complex studied in Ref.\
\onlinecite{Sanda2011}.

Using the coordinate system defined in Fig.\ \ref{fig2} (right panel), the
dipole moments are parametrized according to 
\begin{align}
 {\boldsymbol \mu}_1&=\veco{\mu_1}{z}\\
 {\boldsymbol
\mu}_2&=\veco{\mu_{2}\cos\theta}{x}+\veco{\mu_2\sin\theta\sin\phi}{y}+\veco{\mu_
{2}\sin\theta\cos\phi}{z} . \nonumber
\end{align}
The dipolar coupling is
\be 
J \equiv J(\phi) = \frac{\mu_1\mu_{2}}{R^3} \sin\theta\cos\phi 
\ee
with $R=|\mathbf{R}|$ and $\mathbf{R}=\veco{R}{x}$ and $J_0=\mu_1\mu_2/R^3\simeq
85$ cm$^{-1}$ for the PBDA pair.

\subsection{Slow dynamics of angular movement}

When the dipole moments fluctuate around their orthogonal arrangement, 
the angles evolve stochastically in time and induce a diffusive dynamics 
in a potential. In the simplest model,  the potential is harmonic, and, in more 
detail, we expect a harmonic potential for the angle $\theta$
with equilibrium angle $\theta_0=0$ and free rotation about $\phi$. For a
stochastic time evolution of $\theta$, we assume slow movements described by 
an Ornstein-Uhlenbeck process. This results in an equilibrium density
for $\theta$ of the form 
\be\label{eqrhot}
\rho^{\rm eq}_\theta(\theta) = \frac{1}{\sigma\sqrt{2\pi}}\exp\left[
-\frac{\theta^2}{2\sigma^2} \right]
\ee
and a probability density to observe $\theta$ at time $t$ when the original
angle at initial time $0$ was $\theta'$ as
\bee
{\cal P}_\theta(\theta,\theta',t) &=& \frac{1}{2\sigma^2\pi\sqrt{1-e^{-2Dt}}} \\
&& \times \exp\left[
-\frac{\theta^2+\theta'^2-2\theta\theta'e^{-Dt}}{2\sigma^2(1-e^{-2Dt})} \right].
\nonumber
\eee
Herein, the diffusion constant $D$ is the inverse autocorrelation time of the
fluctuations. Comparing this description with a system-bath approach (as
employed in Ref. \onlinecite{Nalbach2012}), $D^{-1}=\omega_c\lesssim 2500$
cm$^{-1}$. Here, $\omega_c$ is the high-energy cut-off of the environmental
spectral density. The dependence of the chiral 2D spectra on the waiting time
determined below will allow us to extract $D$ from experimental data. The width
of
the angular distribution is $\sigma^2=\langle \theta^2\rangle$. In Ref.
\onlinecite{Nalbach2012} the angular reorganization energy was estimated as 
$\lambda_\theta\simeq 1$ cm$^{-1}$. At high temperatures, we have that 
$\lambda_\theta = \halb\langle J_0^2 \theta^2\rangle/(\kb T)$ with
$J_0=\mu_1\mu_2/R^3\simeq 85$ cm$^{-1}$. Thus, we may estimate at room
temperature for the PBDA pair under consideration the width of $\sigma\simeq
0.24$.

For the angle $\phi$, we assume a homogeneous equilibrium probability density
$\rho^{\rm eq}_\phi(\phi)=1/(2\pi)$. It is reasonable to expect for the dynamics
of $\phi$ a similar form as for the $\theta$-dynamics, but the corresponding
width is $\sigma_\phi\rightarrow\infty$, since free rotation is possible. Since
we are interested on the time scales of the diffusive angular
dynamics and shorter, we model  the dynamics for computational simplicity also
by the form 
\bee
{\cal P}_\phi(\phi,\phi',t) &=& \frac{1}{2\sigma_\phi^2\pi\sqrt{1-e^{-2Dt}}} \\
&& \times \exp\left[
-\frac{\phi^2+\phi'^2-2\phi\phi'e^{-Dt}}{2\sigma_\phi^2(1-e^{-2Dt})} \right] 
\nonumber 
\eee
with $\sigma_\phi=2\pi$.

\subsection{Hamiltonian}

The PBDA pair is a heterodimer. Describing each monomer as a quantum two-level
system, the dimer is described by a Frenkel exciton Hamiltonian 
\be
H = \halb\epsilon_1\sigma_z^{(1)} + \halb\epsilon_2 \sigma_z^{(2)} + J
\sigma_x^{(1)} \sigma_x^{(2)} \, ,
\ee
with the standard Pauli matrices $\sigma_{x,z}^{(j=1,2)}$.
It is readily diagonalized by the transformation
\be
\hat{T} = \exp\left( \ihalb \alpha \sigma_y^{(1)} \sigma_x^{(2)}   + \ihalb\beta
\sigma_x^{(1)} \sigma_y^{(2)} \right) 
\ee
with the angles $\alpha$ and $\beta$ following from 
\[
\tan(\alpha+\beta) = -\frac{J}{\epsilon} \und \tan(\alpha-\beta) =
-\frac{J}{\delta\epsilon}
\]
with $\epsilon=\halb(\epsilon_1+\epsilon_2)$ and
$\delta\epsilon=\halb(\epsilon_1-\epsilon_2)$. For the PBDA heterodimer under
consideration, $\delta\epsilon\simeq 2500$ cm$^{-1}$.
This leads to two independent effective two-level systems described by the
Pauli matrices $\tau_z^{(j=+,-)}$ with the Hamiltonian 
\be
H_d = \hat{T}H\hat{T}^\dagger = \halb E_+\tau_z^{(+)} + \halb E_- \tau_z^{(-)} 
\ee
with $E_{\pm}=\sqrt{J^2+\epsilon^2}\pm\sqrt{J^2+\delta\epsilon^2}$.

\subsection{Light - matter interaction}

The total dipole moment $\hat{\boldsymbol{\mu}} = {\boldsymbol \mu}_1
\sigma_x^{(1)} + {\boldsymbol \mu}_2\sigma_x^{(2)}$ of the dimer is transformed
in the same
way leading to 
\[
\hat{\boldsymbol{\mu}} = {\boldsymbol \mu}_1 \left[ \gamma_1\tau_x^{(+)} -
\gamma_3 \tau_z^{(+)} \tau_x^{(-)} \right] + {\boldsymbol \mu}_2 \left[ \gamma_4
\tau_x^{(-)} - \gamma_2 \tau_x^{(+)} \tau_z^{(-)} \right]
\]
with $\gamma_1=\cos\alpha$, $\gamma_2=\sin\beta$,
$\gamma_3=\sin\alpha$, and $\gamma_4=\cos\beta$.

The two effective two-level systems can be spectroscopically addressed
independently, and thus we focus on the response of the $\tau_\pm$ systems. The
corresponding dipole components are 
\begin{eqnarray}
\hat{\boldsymbol{\mu}}_{+} &= &
\left[ {\boldsymbol \mu}_1 \gamma_1 - {\boldsymbol
\mu}_2  \gamma_2 \tau_z^{(-)} \right] \tau_x^{(+)} \nonumber \, , \\
\hat{\boldsymbol{\mu}}_{-} &= &
\left[ -{\boldsymbol \mu}_1 \gamma_3 \tau_z^{(+)} +
 {\boldsymbol \mu}_2  \gamma_4 \right] \tau_x^{(-)} \nonumber \, .
\end{eqnarray}
Since $\epsilon\gg \delta\epsilon \gtrsim J \simeq \kb T$, we have that 
$\alpha\simeq
-\beta$ and thus $\tan(2\alpha)=-J/\delta\epsilon$. At the same time, 
$\gamma_1=\gamma_4$ and $\gamma_3=-\gamma_2$. Typically, optical spectroscopic
experiments start with the system in the ground state. Hence, we may  
simplify
\bee
\hat{\boldsymbol{\mu}}_{+} &\simeq & \left[ \gamma_1 {\boldsymbol \mu}_1 +
\gamma_2 {\boldsymbol \mu}_2 \right] \tau_x^{(+)} \\
&=&
\bigl[\veco{\gamma_2\mu_{2}\cos\theta}{x}+\veco{\gamma_2\mu_2\sin\theta\sin\phi}
{y} \nonumber \\ &&
\hspace*{3em}+\veco{\gamma_1\mu_1+\gamma_2\mu_{2}\sin\theta\cos\phi}{z}\bigr]
\, ,
\nonumber
\eee
and, similarly,
\bee
\hat{\boldsymbol{\mu}}_{-} &\simeq &
\bigl[-\veco{\gamma_1\mu_{2}\cos\theta}{x}-\veco{\gamma_1\mu_2\sin\theta\sin\phi
}{y} \\ && \hspace*{3em}
+\veco{\gamma_2\mu_1-\gamma_1\mu_{2}\sin\theta\cos\phi}{z} \bigr] . \nonumber
\eee
In the following, we use the notation $\gamma_1\equiv \gamma_1(\phi) =
J(\phi)/{\cal
N}(\phi)$ and $\gamma_2 \equiv \gamma_2(\phi) =
(\sqrt{\delta\epsilon^2+J^2(\phi)}-\delta\epsilon)/{\cal N}(\phi) $ 
%
with ${\cal N}^2 = J^2+(\delta\epsilon-\sqrt{\delta\epsilon^2+J^2})^2$.

Standard multipole expansion of the interaction with the laser electric field
generates
effective magnetic dipoles and electric quadrupoles \cite{Sanda2011}. Those 
typically dominate over the contributions from real magnetic dipoles and
electric
quadrupoles of the two monomers \cite{Somsen1996}. In the following, we neglect 
the latter. Then, we combine the effective magnetic dipole 
moment ${\bf M}_{\pm}$ and the effective
electric
quadrupole tensor ${\bf Q}_{\pm}$ to obtain the tensor 
\begin{equation} \label{thets}
T_{\pm,\alpha\beta}= -i R_{\alpha} \mu_{\mp,\beta}/2 = i Q_{\pm,\alpha\beta}
-\varepsilon_{\alpha\beta\gamma} M_{\pm,\gamma}/k \, ,
\end{equation}
with $R_{\alpha=x,y,z}$ being the components of the distance vector
$\mathbf{R}$. $k$ is the absolute value of the wave vector of the incident
laser field and
where ${Q}_{\pm,\alpha\beta}$ is symmetric and
$\varepsilon_{\alpha\beta\gamma}M_{\pm,\gamma}$ anti-symmetric. We find
\begin{align}
{T}_{-}&=-\frac{iR}{2}
\begin{pmatrix}
\gamma_2\mu_{2}\cos\theta&0&0\\
\gamma_2\mu_2\sin\theta\sin\phi&0&0\\
\gamma_1\mu_1+\gamma_2\mu_{2}\sin\theta\cos\phi&0&0
\end{pmatrix},\\
{T}_{+} &=-\frac{iR}{2}
\begin{pmatrix}
-\gamma_1\mu_{2}\cos\theta&0&0\\
-\gamma_1\mu_2\sin\theta\sin\phi&0&0\\
\gamma_2\mu_1-\gamma_1\mu_{2}\sin\theta\cos\phi&0&0
\end{pmatrix}
\end{align}
and

\begin{figure*}[t]
\epsfig{file=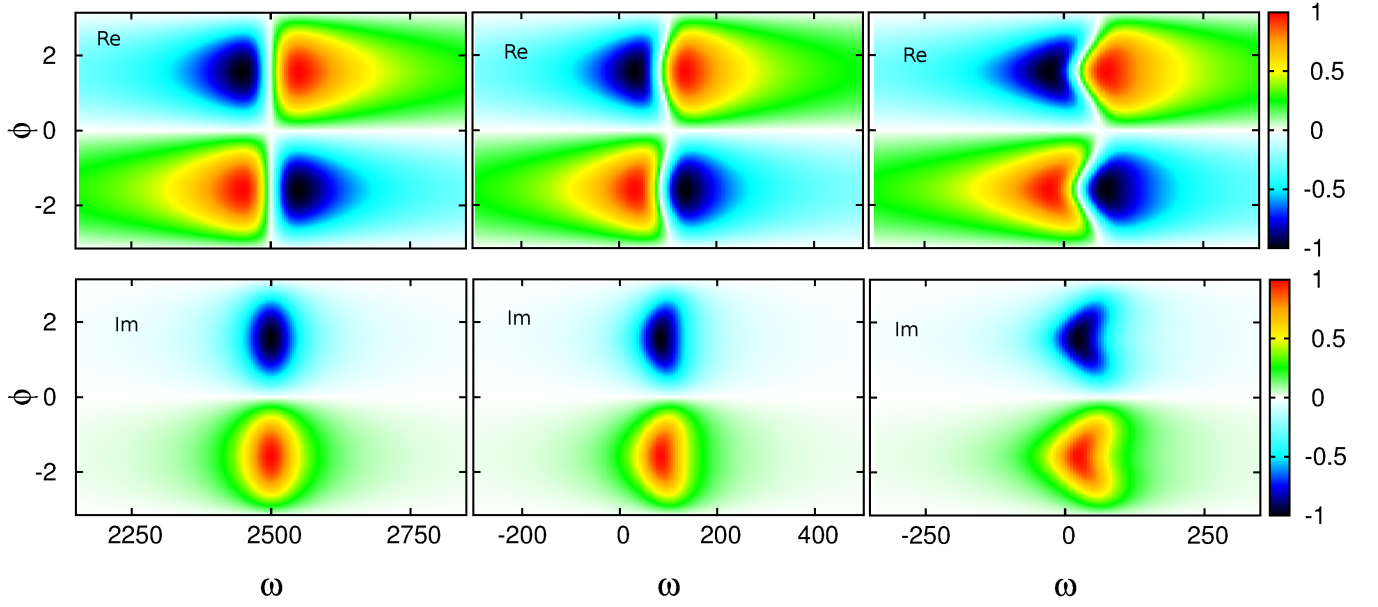,width=18cm}
\caption{\label{fig3} Real (upper row) and the imaginary (lower row) part of the chiral
linear spectrum versus frequency $\omega$ and angle $\phi$ with fixed
$\theta=\pi/4$ for $J=85$cm$^{-1}$, $\Gamma=50$cm$^{-1}$, $\sigma=0.24$ and
$\delta\epsilon=2500$cm$^{-1}$ (left column), $\delta\epsilon=85$cm$^{-1}$ (middle column) and
$25$cm$^{-1}$ (right column).}
\end{figure*}
\begin{widetext}
\begin{align}
{Q}_{-} & = -\frac{R}{4} \cdot \begin{pmatrix}
2\gamma_2\mu_{2}\cos\theta & \gamma_2\mu_2\sin\theta\sin\phi &
\gamma_1\mu_1+\gamma_2\mu_{2}\sin\theta\cos\phi \\
\gamma_2\mu_2\sin\theta\sin\phi & 0 & 0 \\
\gamma_1\mu_1+\gamma_2\mu_{2}\sin\theta\cos\phi & 0 & 0
\end{pmatrix}, \nonumber \\
{Q}_{+} & = -\frac{R}{4} \cdot \begin{pmatrix}
-2\gamma_1\mu_{2}\cos\theta &-\gamma_1\mu_2\sin\theta\sin\phi
&\gamma_2\mu_1-\gamma_1\mu_{2}\sin\theta\cos\phi \\
-\gamma_1\mu_2\sin\theta\sin\phi&0&0\\
\gamma_2\mu_1-\gamma_1\mu_{2}\sin\theta\cos\phi &0&0
\end{pmatrix} \nonumber \, .
\end{align}
\end{widetext}
Moreover, the electric dipole moments assume the form 
\begin{align}
\v{M}_{-}
&=-\frac{ikR}{4}(0,\,\gamma_1\mu_1+\gamma_2\mu_{2}\sin\theta\cos\phi,\,
-\gamma_2\mu_2\sin\theta\sin\phi)^T,  \nonumber \\
\v{M}_{+}
&=-\frac{ikR}{4}(0,\,\gamma_2\mu_1-\gamma_1\mu_{2}\sin\theta\cos\phi,\,
\gamma_1\mu_2\sin\theta\sin\phi)^T. \nonumber
\end{align}

The interaction Hamiltonian of a laser pulse with field
${\mathbf F}(t)\exp(i{\mathbf k}\, {\mathbf R}/2)$ with the dimer in the
rotating wave approximation
\cite{Thimmel99} and to first order in $k=|{\mathbf k}|$ only reads
\bee
H_{\rm int} &=& -\sum_{\alpha=x,y,z} \sum_{j=\pm} F_\alpha \tau^{(j)}_\uparrow
\left\{ \mu_{j,\alpha} + i \sum_{\beta=x,y,z} k_\beta Q_{j,\alpha\beta} \right.
\nonumber\\
&& \left. -\sum_{\beta,\gamma=x,y,z} \varepsilon_{\alpha\beta\gamma} k_\beta
M_{j,\gamma} \right\} + {\rm h.c.} 
\eee
with $\tau^{(j)}_\uparrow=\halb(\tau^{(j)}_x+i\tau^{(j)}_y)$. Moreover,
$F_\alpha$ are the field components of ${\mathbf F}$.

\section{Chiral linear response}

We assume that the two frequencies $E_\pm$ can be probed separately and focus in
the following on $E_+$ which results in a simplified effective field-matter
interaction Hamiltonian $H_{\rm int}=-F_\alpha(t){\cal J}_\alpha\tau_\uparrow +
{\rm h.c.} $ with ${\cal J}_\alpha=\mu_\alpha+ k_\beta T_{\alpha\beta}+\mathcal{O}(k^2)$. 
The tensor elements 
$T_{\alpha\beta}$ follow from Eq.\ (\ref{thets}) after omitting the index $(+)$. 
Moreover, repeated Greek symbols are summed over.

Linear optical response is given by the correlation function
$I_{\alpha\beta}(t)=\langle {\cal J}_\alpha(t){\cal J}_\beta(0) \rangle$.
Experimentally, the response of single molecules is rarely accessible. Instead
ensembles are investigated with each molecule with arbitrary orientation in
space. Thus, orientational averaging (denoted by $\langle\cdot\rangle_\Omega$ in
the following) over the solid angle $\Omega$ 
is performed following standard rules \cite{Andrews1977}. This finally 
results in
\begin{align}
I_{\alpha\beta}(t) &= \frac{1}{3} \delta_{\alpha\beta}
\mean{\mu^*_\gamma(t)\mu_\gamma(0)}_\Omega  \\ 
& \hspace*{-2em}+ \frac{1}{3} \varepsilon_{\alpha\beta\gamma} k_\gamma \bigl[
\mean{M^*_\delta(t) \mu_\delta(0)}_\Omega - \mean{\mu^*_\delta(t)
M_\delta(0)}_\Omega \bigr]. \nonumber
\end{align}
Only the second part is a chiral signal. With light propagating along the 
$x$-direction, i.e.,$k_\alpha=k\delta_{\alpha,x}$, we can observe the chiral component
by measuring 
\bee
I^{\pm}_{yz}(t) &\approx & \frac{2i}{3}\int d\theta \rho^{eq}_\theta(\theta)\int
\frac{d\phi}{2\pi}\cdot \\ 
&& \text{Im}\{M^*_\delta(\theta,\phi)\mu_\delta(\theta,\phi)\} e^{-\Gamma t}e^{-
i E_{\pm}(\theta,\phi) t}\, . \nonumber
\eee
Here, we have assumed that the angles vary only slowly on internal system 
time scales such that we may set $\phi(t)\approx\phi(0)$ and
$\theta(t)\approx\theta(0)$. Herein, $\Gamma$
is the dephasing rate. We assume throughout the paper a dephasing time of $\sim
100$ fs corresponding to $\Gamma=50$ cm$^{-1}$. Fourier transforming the
integrand results in 
\bee
\mathcal{I}^{\pm}_{yz}(\omega,\theta,\phi) &=& \pm\frac{kR\mu_1\mu_2}{6(2\pi)}
\cdot \sin\theta\sin\phi \cdot \rho^\text{eq}_\theta(\theta) \nonumber \\
&& \times
\frac{\omega-E_{\pm}(\theta,\phi)-i\Gamma}{[\omega-E_{\pm}(\theta,
\phi)]^2+\Gamma^2} \, .
\eee
This function is antisymmetric in $\phi$ and $\theta$. Thus, on average the chiral
linear response vanishes. Note that each angle average separately results
already in a vanishing chiral signal.

Fig.\ \ref{fig3} shows the real (upper row) and the imaginary (lower row)
part of the chiral linear signal color coded versus frequency $\omega$ and angle
$\phi$ with fixed $\theta=\pi/4$. We use the parameters as extracted in Ref.\ 
\onlinecite{Nalbach2012}, i.e., $J=85$cm$^{-1}$, $\Gamma=50$cm$^{-1}$ and the standard
deviation $\sigma=0.24$ for the $\theta$-angle fluctuations. We show the results for 
three different values of $\delta\epsilon$, i.e., 
$\delta\epsilon=2500$cm$^{-1}$ (left column), which corresponds to the
PBDA pair experimentally studied \cite{Nalbach2012}, then 
$\delta\epsilon=85$cm$^{-1}$ (middle column) and $25$cm$^{-1}$ (right column). 
For $\phi=0$, 
the complex is achiral and no signal is observed. For finite $\phi$, we observe a
double peak structure with opposite sign for the real part and a single peak in
the imaginary part for varying frequency. For a fixed $\omega$, we observe
also a double peak structure when $\phi$ is varied for
$\delta\epsilon=2500$cm$^{-1}$ (left column). With decreasing $\delta\epsilon$,
however, the peak form (of the imaginary part and of the peak at lower frequency
in the real part) changes towards a {\sl heart} shape exhibiting (for some
$\omega$) two positive peaks followed by two negative ones. We note that the case of 
small $\delta \epsilon$ is closer to a homodimer. Similar results are
expected, when the structure would fluctuate around a chiral equilibrium with
$\phi\not= 0$ and $\theta \not=0$. Those could be used to determine the geometric
structure of such complexes.

\section{2D chiral spectrum}

Next, we focus on nonlinear chiral signals, in particular on 2D chiral spectra. 
We determine the 4-point correlation function \cite{Sanda2011}
\begin{equation}
\mathcal{R}_c=\mathcal{R}^{xx}_{[yz][yz]}+2\mathcal{R}^{zx}_{[xy][yz]},
\end{equation}
which yields a purely chiral 2D signal. Therein, $\mathcal{R}^{zx}$ represents
the response in a set-up where the first two pulses propagate along the
$x$-direction and the third pulse and the detection is along the $z$-direction.
Correspondingly,  $\mathcal{R}^{xx}$ represents a collinear arrangement with all
pulses and detection along the $x$-direction. The subscripts are short-hand
notations for
\bee
\mathcal{R}^{zx}_{[xy][yz]} &=& \mathcal{R}^{zx}_{xyyz} -
\mathcal{R}^{zx}_{yxyz} - \mathcal{R}^{zx}_{xyzy} + \mathcal{R}^{zx}_{yxzy}
\nonumber \\
\mathcal{R}^{xx}_{[yz][yz]} &=& \mathcal{R}^{xx}_{yzyz} -
\mathcal{R}^{xx}_{zyyz} - \mathcal{R}^{xx}_{yzzy} + \mathcal{R}^{xx}_{zyzy}
\nonumber 
\eee
and 
\be 
\mathcal{R}_{\alpha\beta\gamma\delta} =
\mean{\mathcal{J}^*_{\alpha}(\tau_3)\mathcal{J}_\beta(\tau_2)
\mathcal{J}_\gamma(\tau_1)\mathcal{J}^*_\delta(0)}_\Omega
\ee
with  $\mathcal{J}_\alpha=\mu_\alpha+k_\beta T_{\alpha\beta}+\mathcal{O}(k^2)$ as 
defined before. Thus, 
the subscripts in $\mathcal{R}_{\alpha\beta\gamma\delta}$ denote 
the polarization of the pulses, i.e., $\alpha$ for the first pulse, 
$\beta$ and $\gamma$ for the second and third one, and $\delta$ for the detection.
In the following, we focus on the $\tau^{+}$-subsystem. Thereby, we neglect
coherences between the $\tau^{+}$ and $\tau^{-}$ subsystems on the basis of 
assuming the laser pulses
to be spectrally narrower than the energy difference $E_+-E_-$. For this
situation, cross peaks were shown to yield no clear chiral signatures
\cite{Sanda2011}. Hence, we may concentrate on the diagonal peaks.
\begin{figure*}[t]
\epsfig{file=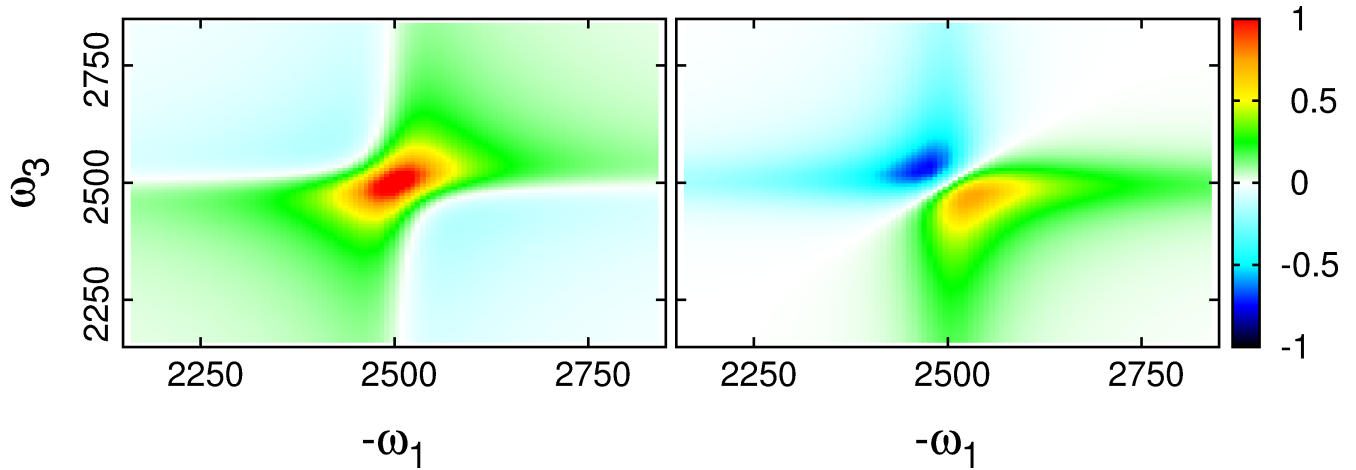,width=18cm}
\caption{\label{fig4} Real (Left) and imaginary part of
$\mathcal{R}_c(\omega_3,t_2,\omega_1)$ for very short waiting times
$Dt_2=10^{-5}$ for the PBDA pair with $J=85$cm$^{-1}$, $\delta\epsilon=2500$cm$^{-1}$ and
$\sigma_\phi=2\pi$ and $\sigma_\theta=0.24$. }
\end{figure*}

Optical signals associated to a chirality exchange result from the 
two-time correlation of
the pseudo-scalar ${\boldsymbol \mu}\cdot\v{M}$. These are weak signals 
and they scale 
as $\propto(kR)^2$. 

We next determine their contribution to the 4-point correlation
function $\mathcal{R}_c$. All contributions which are linear in $kR$ vanish after the
orientational averaging (see Appendix D in Ref.\ \onlinecite{Sanda2011}). Likewise, 
the zero-th order contributions coming from $\mathcal{R}^{zx}_{[xy][yz]}$ also 
vanish for the same reason. Yet, we may consider 
\bee
\mathcal{R}^{xx}_{[yz][yz]} & \propto & \left[
\mean{\mathcal{J}^*_{\alpha}(\tau_3)\mathcal{J}_\beta(\tau_2)
\mathcal{J}_\alpha(\tau_1)\mathcal{J}^*_\beta(0)}_\Omega \right. \nonumber\\
&& \left. - \mean{\mathcal{J}^*_{\alpha}(\tau_3)\mathcal{J}_\beta(\tau_2)
\mathcal{J}_\beta(\tau_1)\mathcal{J}^*_\alpha(0)}_\Omega \right] \, . 
\nonumber
\eee
In case of very fast angular fluctuations, we could average over $\phi$ and
$\theta$ separately at all times $\tau_i$ resulting in a vanishing
$\mathcal{R}^{xx}_{[yz][yz]}$. Typically, however, angular motion is slow
compared to the internal system time scales. Thus, a significant signal strength
is only observed at experimental times $t_1=\tau_1$ and $t_3=\tau_3-\tau_2$,
over which the 2D Fourier transform is later performed, for which $Dt_1, Dt_3
\ll 1$ holds. Thus, the angles do not change during these time intervals. However, the
waiting time $t_2=\tau_2-\tau_1$ is experimentally varied over much longer times
and accordingly angular motion during this time interval has to be taken into
account. This is done by employing the above introduced probability densities to
observe $\theta$ and $\phi$ at time $t_2$ when originally at time $0$ the angles
were $\theta'$ and $\phi'$. Thus, we calculate in detail
%
\bee
& \mean{\mathcal{J}^*_{\alpha}(\tau_3) \mathcal{J}_\beta(\tau_2)
\mathcal{J}_\gamma(\tau_1) \mathcal{J}^*_\delta(0)}_\Omega =
\int d\theta \int d\theta' \int d\phi \int d\phi'  & \nonumber\\
& \times {\cal P}_\phi(\phi,\phi',t_2) {\cal P}_\theta(\theta,\theta',t_2)  \\
& \times \mean{\mathcal{J}^*_{\alpha}(\tau_3,\theta,\phi)
\mathcal{J}_\beta(\tau_2,\theta,\phi) \mathcal{J}_\gamma(\tau_1,\theta',\phi')
\mathcal{J}^*_\delta(0,\theta',\phi')}_\Omega . \nonumber
\eee
%
Thus, for slow angular fluctuations, $\mathcal{R}^{xx}_{[yz][yz]}$ yields no 
zero-th order contribution \cite{Foot1}. Thus, all zero-th and 
first order contributions, i.e., $\propto(kR)^0$ and $\propto(kR)^1$, 
vanish and $\mathcal{R}_c(t_3,t_2,t_1)\propto(kR)^2$. 
Possible contributions to that order from contributions due to the 
light-matter interaction vanish also due to the same arguments \cite{Sanda2011} 
(note that we have neglected these terms above already).

Thus, $\mathcal{R}_c$ is dominated by contributions from the two-time correlations of
the pseudo-scalar ${\boldsymbol \mu}\cdot\v{M}$ of 
order $\propto(kR)^2$. Following the derivation of 
Sanda and Mukamel \cite{Sanda2011}, we find that 
\be
 \mathcal{R}_c(t_3,t_2,t_1) =
-\frac{2k^2}{3}[\mean{X_{\alpha\beta\gamma\gamma\beta\alpha}}_\Omega
+2\mean{X_{\alpha\beta\gamma\beta\alpha\gamma}}_\Omega] \, ,
\ee
with the definition
\bee
 X_{\alpha\beta\gamma\delta\xi\eta} &=&
[-\text{Im}\{T_{\alpha\beta}(\tau_2)\mu_\gamma^*(\tau_2)\}+\text{Im}\{T_{
\gamma\beta}(\tau_2)\mu_\alpha^*(\tau_2)\}] \cdot \nonumber \\
&& \cdot
[-\text{Im}\{T_{\delta\xi}(0)\mu_\eta^*(0)\}+\text{Im}\{T_{\eta\xi}
(0)\mu_\delta^*(0)\}]. \nonumber
\eee
For our PBDA dimer, we have that 
\bee
\langle X_{\alpha\beta\gamma\gamma\beta\alpha}\rangle &=&
-2\left(\frac{R}{2}\right)^2\mu_1^2\mu_2^2 \cdot \nonumber \\ && \cdot
\langle\cos\theta_2\cos\theta_0
+\sin\theta_2\sin\phi_2\sin\theta_0\sin\phi_0\rangle_{\Omega} \nonumber\\
\langle X_{\alpha\beta\gamma\beta\alpha\gamma}\rangle &=&
\left(\frac{R}{2}\right)^2\mu_1^2\mu_2^2\langle\cos\theta_2\cos\theta_0\rangle_{
\Omega}  \nonumber
\eee
with $\theta_i\equiv\theta(\tau_i)$ and $\phi_i$ likewise.

Next, we observe that the energy $E_+\equiv E_+(\theta,\phi)$ is a function of
the angles $\theta$ and $\phi$ and that the primed energy means that the 
energy depends on the primed angles. With $t_1=\tau_1$, $t_2=\tau_2-\tau_1$ and
$t_3=\tau_3-\tau_2$, we obtain after a Fourier transformation with respect to
the times $t_1$ and $t_3$ that 
%
\bee
\mathcal{R}_c(\omega_3,t_2,\omega_1) &=& \frac{(kR)^2}{3}\mu_1^2\mu_2^2 \iiiint
d\theta d\theta' d\phi d\phi' \cdot \hspace*{5mm} \\
&& \hspace*{-1cm}\times {\cal P}_\phi(\phi,\phi',t_2) {\cal
P}_\theta(\theta,\theta',t_2) r(\theta, \phi, \theta', \phi') \nonumber 
\eee
with 
\bee
r(\theta, \phi, \theta', \phi') &=& \sin\theta\sin\phi\sin\theta'\sin\phi' \\
&& \times \frac{\Gamma+i(\omega_3-E_+)}{\Gamma^2+(\omega_3-E_+)^2}
\frac{\Gamma+i(\omega_1+E_+')}{\Gamma^2+(\omega_1+E_+')^2}  . \nonumber
\eee

Fig.\ \ref{fig4} shows the 2D chiral signal
$\mathcal{R}_c(\omega_3,t_2,\omega_1)$ for a very short waiting time
$Dt_2=10^{-5}$. We employ the parameters of the PBDA pair, i.e.,
$J=85$cm$^{-1}$, $\delta\epsilon=2500$cm$^{-1}$ and $\sigma_\phi=2\pi$ and
$\sigma_\theta=0.24$. The left (right) panel depicts the real (imaginary) part.
Similar peak shapes are observed for smaller $\delta\epsilon$ and for longer
waiting times. The peak shape reflects the assumption of a Markovian dephasing
dynamics which is inherent to our assumption of the dephasing taken 
in the form of an exponential decay with a fixed dephasing time $\Gamma$. 
The integrand $r(\theta, \phi, \theta', \phi')$ determines which angular 
conformation dominantly contributes to the signal. 
As expected, these are the chiral conformations ${\boldsymbol \mu}_2 
\perp {\boldsymbol \mu}_1\parallel {\bold n} \perp {\boldsymbol \mu}_2$.

\begin{figure}[t]
\epsfig{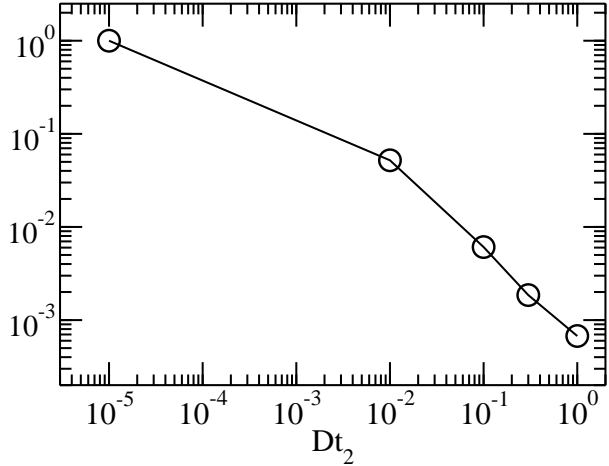}
\caption{\label{fig5} Maximal amplitude of the real part 
$R(t_2)=\mathrm{max}_{\omega_1,\omega_3} \mathcal{R}_c(\omega_3,t_2,\omega_1)$ scaled to 
$R_0=R(t_2=10^{-5}D^{-1})$ versus waiting times $Dt_2$ for
$J=85$cm$^{-1}$, $\delta\epsilon=2500$cm$^{-1}$ and $\sigma_\phi=2\pi$ and
$\sigma_\theta=0.24$. }
\end{figure}

We expect that the overall strength of the nonlinear chiral signal 
rapidly decreases with increasing waiting times on a time scale 
determined by the autocorrelation time of the angular fluctuations. 
This is depicted in Fig.\ \ref{fig5} which shows the maximal amplitude 
$R(t_2)=\mathrm{max}_{\omega_1,\omega_3} \mathcal{R}_c(\omega_3,t_2,\omega_1)$ 
scaled to $R_0=R(t_2=10^{-5}D^{-1})$. Nonlinear chiral signals are in general 
 weak since typically $(kR)^2\sim 10^{-6}$. The fact that only non-vanishing 
fluctuations lead in the present case to a finite contribution to the signal 
only reduces the signal strength by another factor $R_0\simeq 0.08$.
In turn, by determining the peak maximum of
the chiral 2D signal $\mathcal{R}_c(\omega_3,t_2,\omega_1)$ for various waiting
times, one can experimentally measure the autocorrelation time of the angular
fluctuations.

\section{Conclusions}

We have determined the linear and 2D optical chiral spectra for a dimer system
whose dipole moments are orthogonal with a connecting vector orthogonal to one
of the dipoles. In its equilibrium configuration, the dimer is achiral and no
chiral signal is expected. At the same time, the dipolar coupling vanishes and no
F\"orster-type energy transfer arises. Geometrical fluctuations, however,
result in finite dipolar couplings, causing rather fast energy transfer which has been
experimentally observed. We show that to assign the fast energy transfer
unambiguously to angular fluctuations around an orthogonal equilibrium
configuration, chiral signals can be used. As long as the dimer configuration
fluctuates around an orthogonal equilibrium configuration, the linear chiral
spectrum vanishes, but is non-zero for (fluctuating) non-orthogonal
configurations. The nonlinear 2D spectrum vanishes in the equilibrium
configuration but is non-zero when angular fluctuations are present as long as
the waiting time is short compared to the autocorrelation time of the
fluctuations. This, in turn, can be used to determine the autocorrelation 
times experimentally. Hence, our approach may also be used to reveal 
correlation times of fluctuations which are otherwise hardly accessible experimentally.

\section{Acknowledgements}
We acknowledge support by the DFG Sonderforschungsbereich 925 
``Light-induced dynamics and control of correlated quantum systems'' 
(Project C8) and by the DFG excellence cluster ``The Hamburg Center for
Ultrafast Imaging''. S.M. gratefully acknowledges the support of the NSF through
Grant No. CHE 1361516, NIH Grant No. GM-59230, and the Chemical Sciences,
Geosciences and Biosciences Division, Office of Basic Energy Sciences, Office of
Science, US Department of Energy.


\begin{thebibliography}{99}
\bibitem{Foot1} To treat the case where the angular motion happens on the same time
scale as the internal system time scales $E_\pm^{-1}$, a full quantum mechanical
treatment of the angular dynamics would be required.
\end{thebibliography}
\end{document}